\documentclass[pr,twocolumn,showpacs,superscriptaddress]{revtex4}
\usepackage{graphicx}
\usepackage{dcolumn}
\usepackage{amsmath,amsbsy,amssymb,graphics}
\usepackage{bm}
\usepackage{mathrsfs,color}

\begin{document}

\preprint{APS/123-QED}

\title{
On the limitations of cRPA downfolding
}

\author{Carsten Honerkamp}
\affiliation{Institute for Theoretical Solid State Physics,  RWTH Aachen University, and JARA Fundamentals of Future Information Technology, Germany}
\author{Hiroshi Shinaoka}
\affiliation{$^2$ Department of Physics, Saitama University, Saitama 338-8570, Japan}
\author{Fakher F. Assaad}
\affiliation{Institut f\"ur Theoretische Physik und Astrophysik, Universit\"at W\"urzburg,
Am Hubland, D-97074 W\"urzburg, Germany}
\author{Philipp Werner}
\affiliation{Department of Physics, University of Fribourg, 1700 Fribourg, Switzerland}

\date{\today}

\hyphenation{instabi-li-ty compa-tible}

\begin{abstract}
We  check the accuracy of the constrained random phase approximation (cRPA) downfolding scheme by considering one-dimensional two- and three-orbital Hubbard models with a target band at the Fermi level and one or two screening bands away from the Fermi level. Using numerically exact quantum Monte Carlo simulations of the full and downfolded model we demonstrate that depending on filling the effective interaction in the low-energy theory is  either barely screened, or antiscreened, in contrast to the cRPA prediction. This observation is explained by a functional renormalization group analysis which shows that the cRPA contribution to the screening is to a large extent cancelled by other diagrams in the direct particle-hole channel. We comment on the implications of this finding for the ab-initio estimation of interaction parameters in low-energy descriptions of solids. 
\end{abstract}

\maketitle

\section{Introduction}

Downfolding describes a process by which a low-energy effective theory for some material is constructed by systematic elimination of high-energy degrees of freedom. The first step in an ab-initio simulation of a material is typically a bandstructure calculation based on the local density approximation \cite{Sham1965} or single-shot $G_0W_0$ \cite{Hedin1965} in a large energy window, while the goal is to compute the electronic structure in a relatively small energy window around the Fermi level. A systematic procedure to derive the interaction parameters for the low-energy theory in this window is the constrained random phase approximation (constrained RPA or cRPA) \cite{Aryasetiawan2004}. The idea is to calculate the screening contribution of all processes involving high-energy bands using the random phase approximation and based on this evaluate the partially screened effective interaction for the low-energy theory. This interaction is in general nonlocal and frequency dependent \cite{Werner2016}. A more systematic approach, which takes into account also non-RPA contributions to the polarization function, is based on functional renormalization group (fRG) calculations \cite{Honercamp2012,Ginza2015,Honerkamp2018}. Here, the integration over the high-energy bands includes all one-loop diagrams, and thus also incorporates particle-particle screening and magnetic fluctuations. By restricting the cfRG to just one class of diagrams, the cRPA can be recovered as a particular approximation. 

While cRPA tools are nowadays built into several electronic structure codes \cite{Amadon2014, fleur, Merzuk2015}, and the method is widely used to estimate the interaction parameters for low-energy effective theories, the cfRG method has so far mainly been applied to toy models. There, depending on the model, the effective target-band interactions showed clear deviations from the cRPA results. For instance, cases with antiscreening of the effective repulsion due to magnetic fluctuations were identified \cite{Ginza2015}. The inclusion of additional channels causes a more complex wavevector- and frequency-dependence, which is most clearly visible in the non-local effective interactions that in the cfRG can, for instance, contain antiferromagnetic interactions that cannot be obtained by cRPA \cite{Honerkamp2018}. 

What has been missing up to now are rigorous tests of the cRPA and cfRG schemes. A possible strategy is to check the accuracy of the downfolded interaction in simple but well-controlled model set-ups. For the case of cRPA, the downfolding from a three-dimensional three-band Hubbard model to a single-band Hubbard model \cite{Shinaoka2015} has been tested using extended dynamical mean field theory (EDMFT) \cite{Sun2002}. This formalism allows to approximately describe the effect of dynamical and nonlocal interactions, and thus to compare the properties of the downfolded single-band model to those of the original three-band model. This study revealed rather severe deviations, and in particular showed that the cRPA interaction tends to overestimate the screening effect from the two screening bands \cite{Shinaoka2015}. Without interaction between the screening bands, the EDMFT result for the downfolded model with {\it bare} interactions produced a significantly better agreement with the EDMFT result of the full model than the calculation employing the cRPA interaction. Given the fundamental importance of accurate downfolding schemes for the ab-initio simulation of materials, it is important to understand the deficiencies of the cRPA approach, and to perform additional model studies. 

A limitation of the approach in Ref.~\onlinecite{Shinaoka2015} was the use of EDMFT, which provides only an approximate solution of the lattice problem. To elimintate the resulting uncertainty, we consider here a downfolding from a one-dimensional three-band model to an effective one-dimensional single-band model. Both models can be solved exactly using lattice Quantum Monte Carlo (QMC) simulations \cite{Assaad2007}, which allows an accurate assessment of the quality of the cRPA downfolding. To gain insights into the failures of the cRPA construction in this particular model setup we perform a cfRG analysis and identify the relevant diagrammatic contributions to the polarization function. The overscreening in the cRPA formalism can thus be explained by missing cancellation effects between cRPA and non-cRPA diagrams. 

The paper is organized as follows: in Section~\ref{sec:model} we introduce the model Hamiltonians and briefly explain the cRPA and cfRG downfolding schemes, as well as the QMC simulation of the full and downfolded lattice models. Section~\ref{sec:results} presents the QMC based benchmark calculations for the three-band model and cRPA interaction, and an analysis of the result based on cfRG. A discussion and conclusions are presented in Sec.~\ref{sec:discussion}.

\section{Model and Methods}
\label{sec:model}

\subsection{Model Hamiltonian}

We consider a two- or three-orbital Hubbard model on a one-dimensional chain with orbital-diagonal transfer $t=1$ between nearest-neighbor sites. The Hamiltonian of the {\em full model}  is given by
\begin{eqnarray}
  \mathcal{H} &=& - \sum_{\langle i,j\rangle}\sum_{\alpha\sigma} \hat{c}^\dagger_{i\alpha\sigma}\hat{c}_{j\alpha\sigma} + \sum_i \sum_\alpha (E_\alpha+E^\mathrm{dc}_\alpha-\mu) \hat{n}_{i\alpha} \nonumber \\
  &&-t^\prime \sum_i \sum_{\sigma}\sum_{\beta\neq 2} \left(\hat{c}^\dagger_{i2\sigma}\hat{c}_{i\beta\sigma} + \hat{c}^\dagger_{i\beta\sigma}\hat{c}_{i2\sigma} \right)\nonumber\\
  &&+\sum_{i\alpha} U_\alpha \hat{n}_{i\alpha\uparrow}\hat{n}_{i\alpha\downarrow},\label{eq:Ham}
\end{eqnarray}
where $i$ and $j$ are site indices, while $\alpha$ and $\beta$ are orbital indices taking the values 1 and 2 for the two-band model, with band 1 being the target band, or 1, 2, 3, with band 2 as target band for the three-band model. 
The chemical potential is $\mu$.

\begin{figure}
 \centering\includegraphics[width=.45\textwidth,clip]{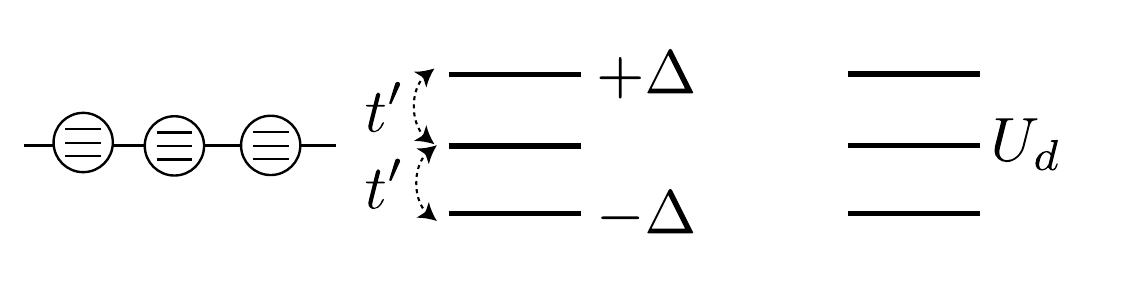}\\
 \includegraphics[angle=0,width=\columnwidth]{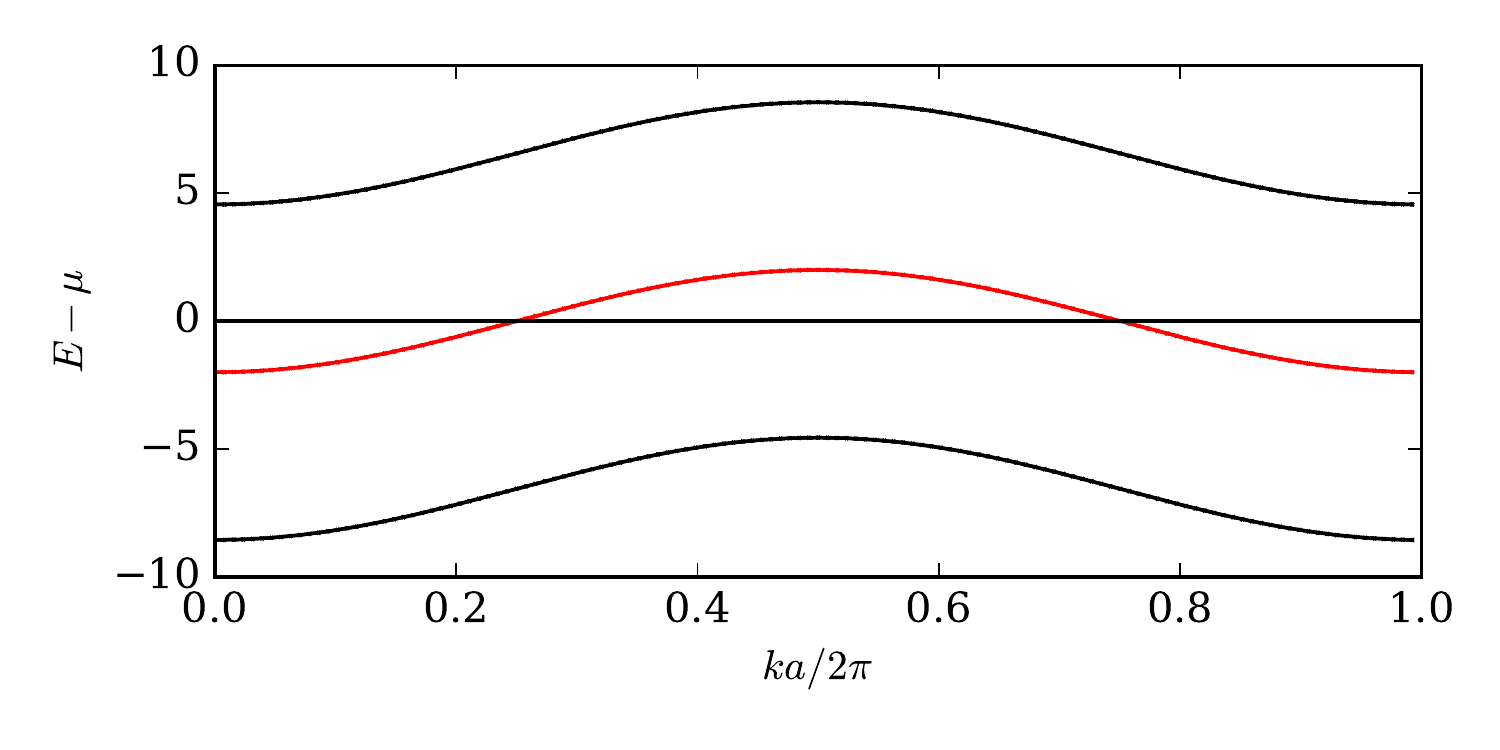}
 \caption{
 Top panels: Schematic visualization of the three-orbital model on a one-dimensional chain. The three orbital levels are split by orbital-dependent on-site energies. The hopping along the chain is orbital-diagonal. We include an orbital-offdiagonal transfer $t^\prime $ between the central orbital and the upper and lower orbital, but the latter two  are not connected by any matrix element. The on-site repulsion for the central orbital is denoted by $U$, while the highest and lowest orbitals are non-interacting. 
 Bottom panel: non-interacting bandstructure for $\Delta=5$, with the target band (red) near the Fermi level and two screening bands centered at $\pm\Delta$. In the two-band case, the upper band is missing.
 }
\label{fig:model1}
\end{figure}

We consider only onsite density-density interactions \cite{footnote1}. The on-site repulsion $U_\alpha$ 
is taken to be orbital-dependent.
In the two-orbital cases below, we take $U_1=U_2=4$, while for the three-band case we choose $U_\alpha=0,~U,~0$ for $\alpha=1,2,3$, respectively ($U>0$). This choice is reasonable,  because screening bands are usually less correlated than target bands in real materials (see illustration in Fig.~\ref{fig:model1}). 
The orbital potentials $E_\alpha$ are given by $-\Delta$, $0$, $\Delta$ for $\alpha=1,2,3$ in the three-orbital case,
and $E_\alpha = 0$, $\Delta$ for $\alpha=1,2$ in the two-orbital case. 
$\Delta>0$ produces (direct) gaps between the target- and screening-band manifolds.

The cRPA and cfRG downfolding is performed in the noninteracting bandstructure with $E^\mathrm{dc}_\alpha=0$, which means that self-energy corrections to the bandstructure are not taken into account.  
To make the model particle-hole symmetric in the QMC simulations,
we set $E^\mathrm{dc}_\alpha=0$, $-U/2$, 0 for $\alpha=1,2,3$. 
At first sight, this appears to be a change of the model, but at half-filling and in perturbation theory, this shift exactly cancels the first-order Hartree self-energy. Hence studying the model in cRPA or cfRG with $E^\mathrm{dc}_{\alpha}= 0$,  as done here, is at half-filling equivalent to absorbing the Hartree shift into the bare bands.     

The parameter $t'$ controls the onsite inter-orbital hopping between the central orbital 2 and the two outer orbitals 1 and 3. There is no direct hopping between orbitals 1 and 3.  We set $t^\prime$ to $1$ and 3 in the two- and three-orbital case, respectively. 

\begin{figure}
	\centering\includegraphics[width=.5\textwidth,clip]{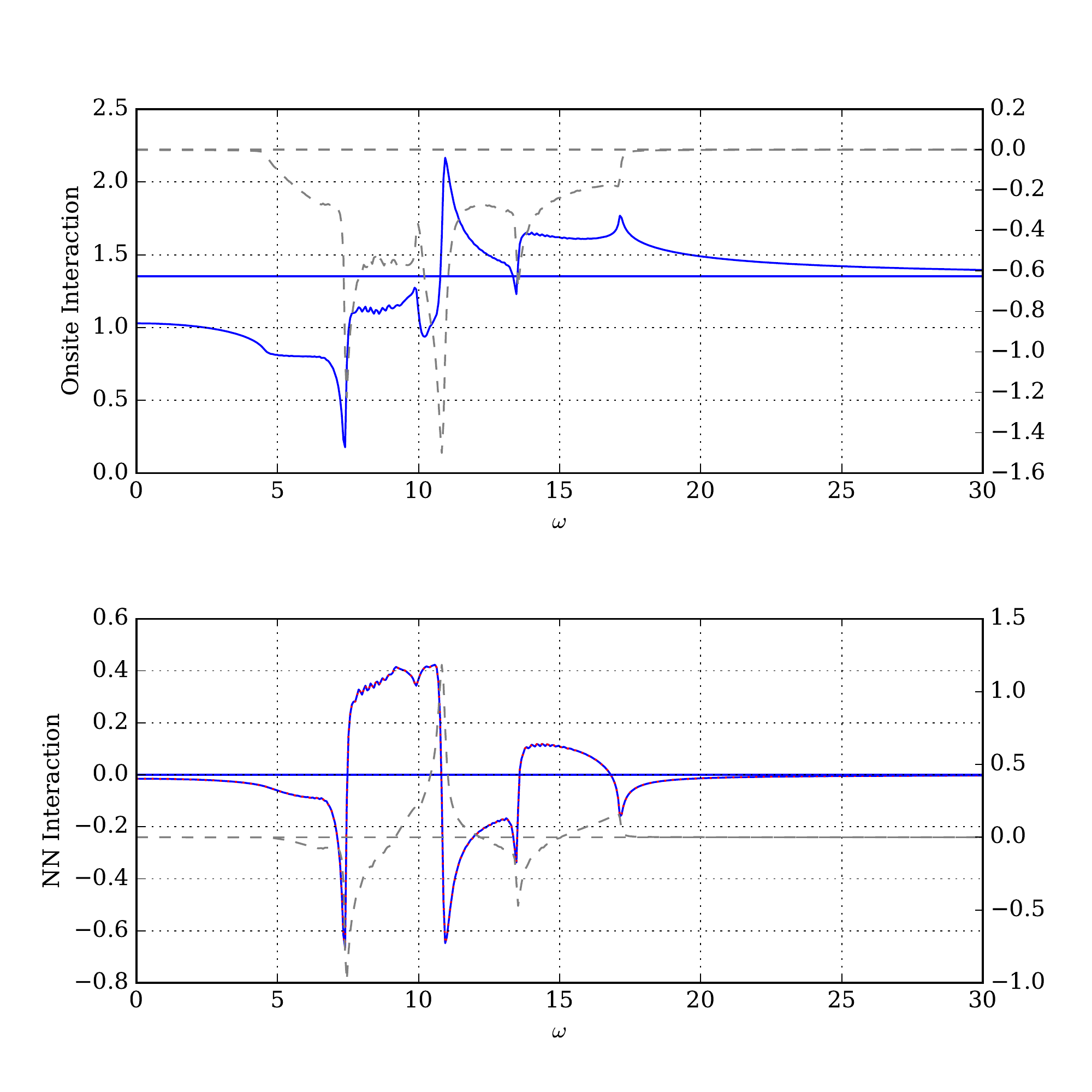}
	\caption{
		Effective onsite and nearest-neighbor (NN) interactions computed by cRPA for the three-orbital model with $U = 4$ and $\Delta=5$ (same as Fig.~\ref{fig:model1}).
		The real and imaginary parts are shown by solid and broken lines, respectively.
		The left/right vertical axis is for the real/imaginary part of the screened interaction.
	}
	\label{fig:model1-cRPA}
\end{figure}

\subsection{cRPA procedure}

In the cRPA approach we downfold the full model to a single-band model by summing up the constrained one-loop polarization, in which at least one internal line has to be in the high-energy bands \cite{Aryasetiawan2004}, using the random phase approximation. Then we  project the obtained partially screened interaction onto the (Wannier functions of the) target band. Only density-density interactions are kept in our cRPA model.
Also the kinetic term is evaluated for the Wannier functions of the target band.
In the present set-up, with possible self-energy corrections unaccounted for, the downfolded model exhibits the same  nearest-neighbor hopping dispersion as the bare band.

The action of the downfolded single-band model reads
\begin{eqnarray}
	\tilde S &=&   \tilde S_0  + \int_0^\beta d\tau \sum_i \tilde U n_{i,\uparrow}(\tau) n_{i,\downarrow}(\tau) \nonumber\\
	&&+ \int_{0}^{\beta}  d \tau   \sum_{i<j} \tilde V_{ij}   n_i (\tau)   n_j (\tau)  \nonumber\\
	&&+ \int_{0}^{\beta}  d \tau  d \tau'  \sum_{i\le j} \tilde W_{ij}(\tau-\tau')  n_{i} (\tau)    n_{j} (\tau'),
	\label{eq:downfolded}
\end{eqnarray}
where $n=n_\uparrow+n_\downarrow$ and  $\tilde S_0$ denotes the free system projected onto the target band. 
As an illustration, we show in Fig.~\ref{fig:model1-cRPA} the real-frequency dependence of the  on-site and nearest-neighbor interaction of the cRPA-downfolded model for the parameters $U=4$ and $\Delta=5$.
The high-frequency limit of the on-site interaction equals the bare target-band interaction. Its value for these parameters is around 1.35 in units of the intra-orbital hopping , which is smaller than $U$ on the central orbital.
This reduction is due to the `spreading' of the central, interacting  orbital, which is quite strongly hybridized with two non-interacting orbitals,  in the three bands of the dispersion, as encoded in the orbital-to-band transformation. At low frequencies, the cRPA screens this value down to $\sim 1.03$, i.e. by roughly 24$\%$. The nearest-neighbor interaction remains small, especially in the low-frequency range. Hence the effective cRPA interaction essentially corresponds to a one-band model with a reduced local repulsion.  

In addition to the so-obtained {\em cRPA-screened low-energy model} we also construct the {\em unscreened low-energy model} with the same kinetic term as the cRPA-screened model but with the {\it bare} Coulomb interaction projected onto the target band. This yields the above-mentioned value of 1.35 for the onsite repulsion of the three-orbital model for the chosen parameters.

\subsection{cfRG procedure}
Next we explain how we use the constrained functional renormalization group (cfRG) to construct a {\em cfRG-screened low-energy model}, again only with one target band in our case.
 
\begin{figure}[ht] 
\includegraphics[width=.98\columnwidth]{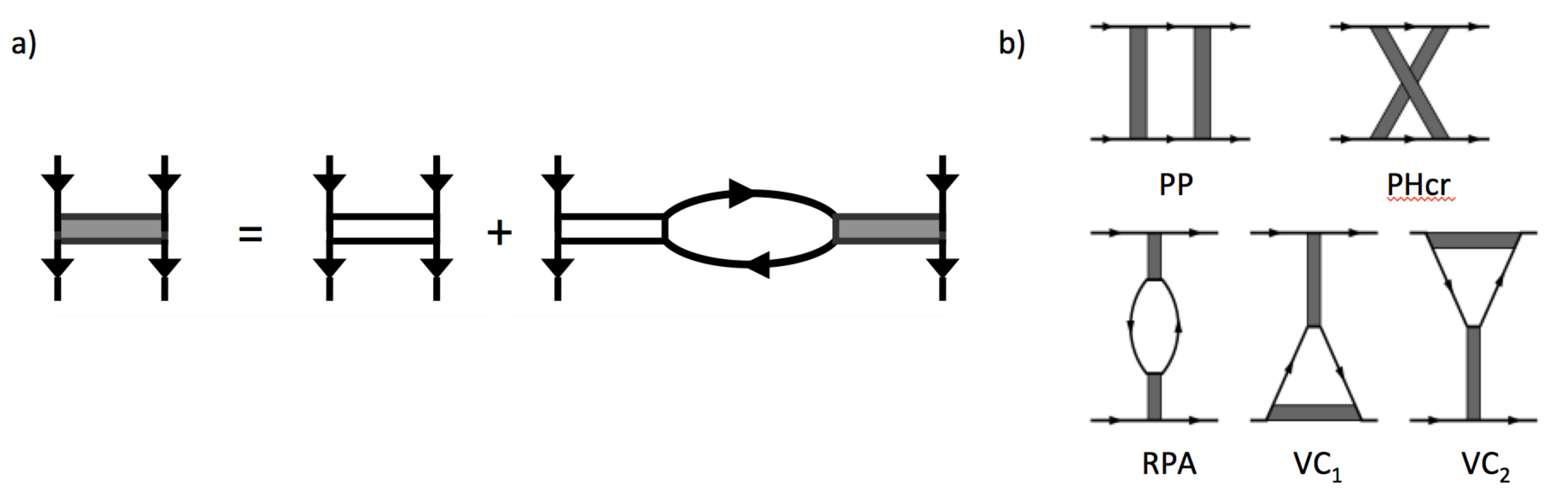}
 \caption{a) Bethe-Salpeter equation for the effective  interaction in cRPA. The empty rectangles denote the bare  interaction and the full ones the screened interaction. The spin projection is conserved along the short edge of the rectangle (assuming spin-rotational invariance).  b) Five one-loop diagrams on the right hand side of the fRG flow equation for the interaction vertex. cfRG takes into account all these diagrams. The internal lines carry cutoff functions and get differentiated with respect to the RG flow parameter in the flow equation. Only taking into account the diagram denoted as RPA is equivalent to the cRPA of a).}
\label{fig:diagrams}
\end{figure}

The cfRG calculation also starts with the full non-interacting band structure. Then, a cutoff function is associated with the bare Green's function. This allows one to continuously integrate over the high-energy bands, leaving the target bands untouched. 
The continuous integration gives rise to the RG flow of the effective interactions. Due to the loop corrections, the effective interactions can acquire a pronounced wavevector- and frequency dependence. In this paper, we use the intraorbital onsite and instantaneous  bilinear interaction approximation of Ref.~\onlinecite{Honerkamp2018}. 
This means that the effective interaction is represented in terms of fermion charge, spin and pair bilinears that reside in the same orbital at the same lattice site and the same Matsubara time, which can interact with bilinears of the same type at different sites.  In this approximation, the coupling between these bilinears depends on one collective wavevector and one bosonic Matsubara frequency.
More precisely, we write the total two-fermion interaction, which is a function of three wavevectors and frequencies labeled by $k_i=(\vec{k}_i,\omega_i )$, as the sum of three channels,
\begin{eqnarray}
V^\Lambda_{o_1o_2o_3o_4} (k_1,k_2,k_3) &=& P_{o_1o_3}^{\Lambda} (k_1+k_2) +  D_{o_1o_2}^{\Lambda} (k_3-k_1)\nonumber \\ && +  C_{o_1o_2}^{\Lambda} (k_3-k_2) \, .
\label{viobi} 
\end{eqnarray}
The indices 1 and 2 belong to the incoming particles, while 3 and 4 denote the outgoing particles.
$\Lambda$ is the flow parameter. The bare density-density interaction at the initial scale $\Lambda_0$ can be used as initial condition for  $D_{o_1o_2}^{\Lambda} (k_3-k_1)$, while potential Hund's rule terms give rise to nonzero initial conditions for the other two channels.
The interaction (\ref{viobi}) does not carry spin indices and it is understood that the spin projection of particles 1 and 3 as well as those of 2 and 4 are pairwise the same, in order to fulfill spin rotation symmetry. From $V^\Lambda_{o_1o_2o_3,o_4} (k_1,k_2,k_3)$, the full two-particle vertex for any allowed spin combination can be reconstructed.
Upon integration over the high-energy modes, the coupling function $ P_{o_1o_3}^{\Lambda} (k_1+k_2)$ flows with the particle-particle diagram (PP) on the right hand side of Fig.~\ref{fig:diagrams},  $D_{o_1o_2}^{\Lambda} (k_3-k_1)$ flows with the direct particle-hole terms denotde by RPA, VC1 and VC2, while  $C_{o_1o_2}^{\Lambda} (k_3-k_2)$ flows with the crossed particle-hole diagram PHcr. At the two vertices appearing in these diagrams, one has to insert the full flowing vertex (\ref{viobi}). For details of how the insertion of the coupling functions is handled effciently, see Ref.~\onlinecite{Honerkamp2018}. This insertion establishes the coupling between the different coupling functions. If we drop all diagrams except the RPA diagram, solving the fRG differential equations amounts to solving the Bethe-Salpeter equation for  $D_{o_1o_2}^{\Lambda} (k_3-k_1)$, i.e. the cRPA. If we keep all diagrams, we call the procedure cfRG.
The use of the Wick-ordered flow \cite{Salmhofer1998,Honercamp2012} permits to also capture the mixed one-loop diagrams that have one internal target band line. At the end of the flow, when only the target bands are left, the effective target-band interaction can be Fourier transformed into real space, leading to distance-dependent effective interactions. They can also be expressed in terms of charge and spin interactions. Via the comparison with the bare values effective screening functions for the downfolding process can be computed. For more details we refer to Ref. \onlinecite{Honerkamp2018}. 

We note that in the cfRG used here we ignore self-energy corrections in all bands.  This approximation is mainly made because in this first comparison we want to focus on the corrections to the effective interactions and disentangle this from other effects. 
Thus  the current cfRG misses in particular target band renormalizations in the downfolding, which might add to observed changes in the effective interactions.   
In principle, the fRG allows one to include self-energy corrections, as recently demonstrated for the one-band Hubbard model in Refs. \onlinecite{vilardi,tagliavini}.   However this has not been explored yet in a multi-orbital- or cfRG-context and is left for future work. 
Note also that the models studied here have an overall bandwidth of up to $16t$ with strong hybridization such that they are, viewed on the bare energy scale, not strongly interacting. Thus, the self-energy corrections due to the high-energy bands should be only Hartree-type (see the discussion of $E_{\mathrm{dc}}$ above) or otherwise minor quantitative corrections.  

\subsection{QMC simulations}

To solve  the  full  and downfolded model we use a general implementation of the CT-INT algorithm \cite{Gull2011} that allows for arbitrary range density-density interactions in space and time.  The general form of the action is 
\begin{eqnarray}
	   S &=&   S_0  +   \int_{0}^{\beta }  d \tau   \sum_{x} U(x)   n_{x, \uparrow} (\tau)  n_{x,\downarrow} (\tau)    \nonumber   \\  
	          & &    + \frac{1}{2}  \int_{0}^{\beta }  d \tau   \sum_{x \neq y} V(x,y)   n_{x} (\tau)   n_{y} (\tau)     \\
	           & &    + \frac{1}{2} \int_{0}^{\beta }  d \tau  d \tau'  \sum_{x , y} W(x,y, \tau-\tau')  n_{x} (\tau)    n_{y} (\tau'),  \nonumber 
\end{eqnarray}
where $ S_0 = \int_{0}^{\beta }  d \tau    \sum_{x,y,\sigma}c^{\dagger}_{x,\sigma}(\tau)  \left( \frac{\partial}{\partial \tau } \delta_{x,y}  -  T_{x,y} \right)   c^{\phantom\dagger}_{y,\sigma}(\tau) $ is the action of the free system  with band structure given by $T_{x,y}$. This form of the action is appropriate to describe both the original multiband model (\ref{eq:Ham}) (with $V=0$ and $W=0$) and the downfolded model (\ref{eq:downfolded}). In the above notation $x$ is a composite index which  combines site and orbital indices. 
These models are solved at inverse temperature  $\beta=8$ and on a 10-site  chain with periodic boundary conditions.
We determine the chemical potential of the full model such that the total electron number is $n_\mathrm{target}+2$, where $n_\mathrm{target}$ is the electron number of the corresponding downfolded model. 

We use the approach described in Ref.~\onlinecite{Assaad2007}  for the retarded interactions and that of Ref.~\onlinecite{Hohenadler2012} for  equal time correlations. 

\section{Results}
\label{sec:results}

\subsection{QMC tests of the cRPA interaction}

\begin{figure}[b]
 \centering\includegraphics[width=.235\textwidth,clip]{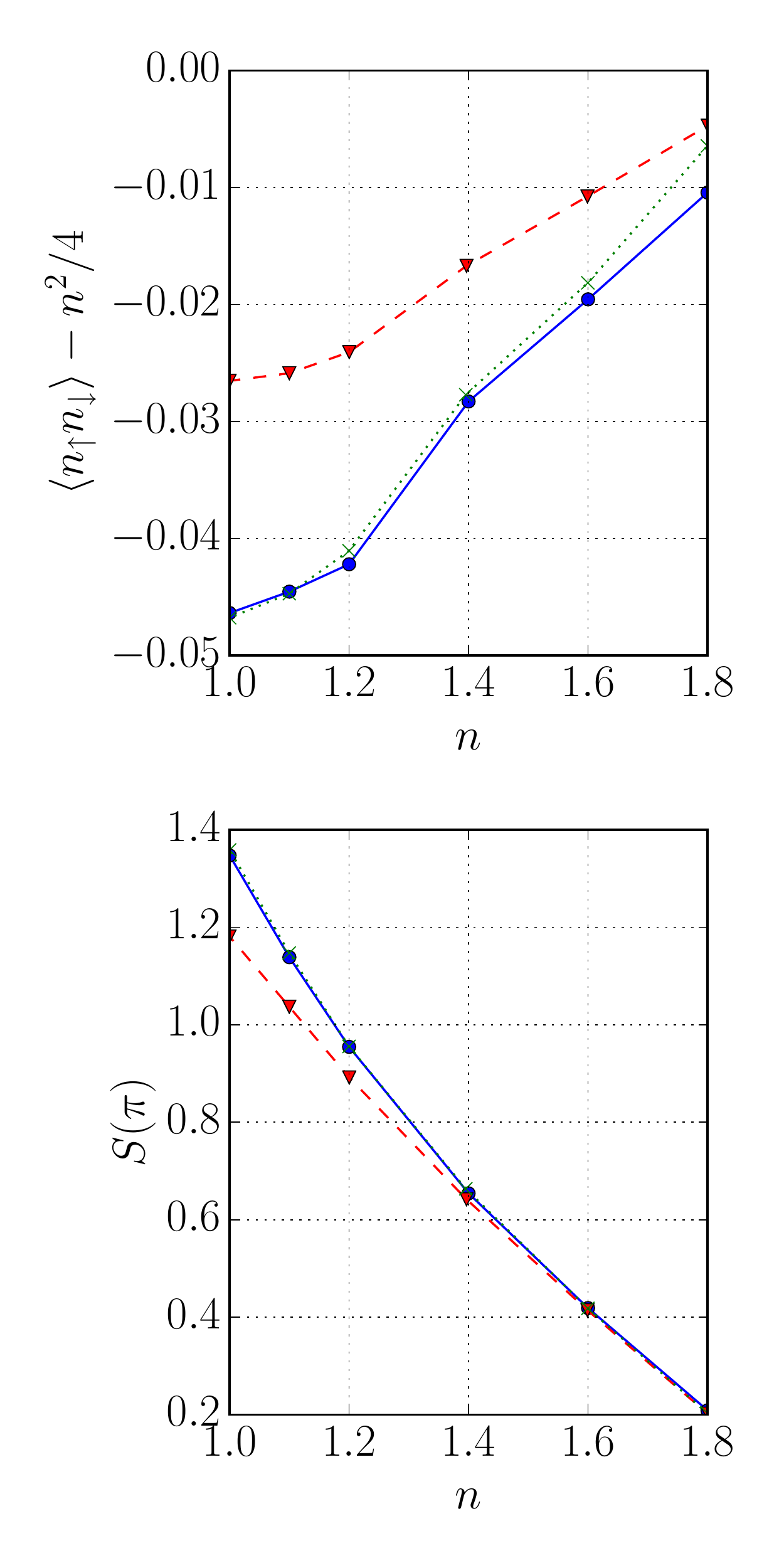}
 \centering\includegraphics[width=.235\textwidth,clip]{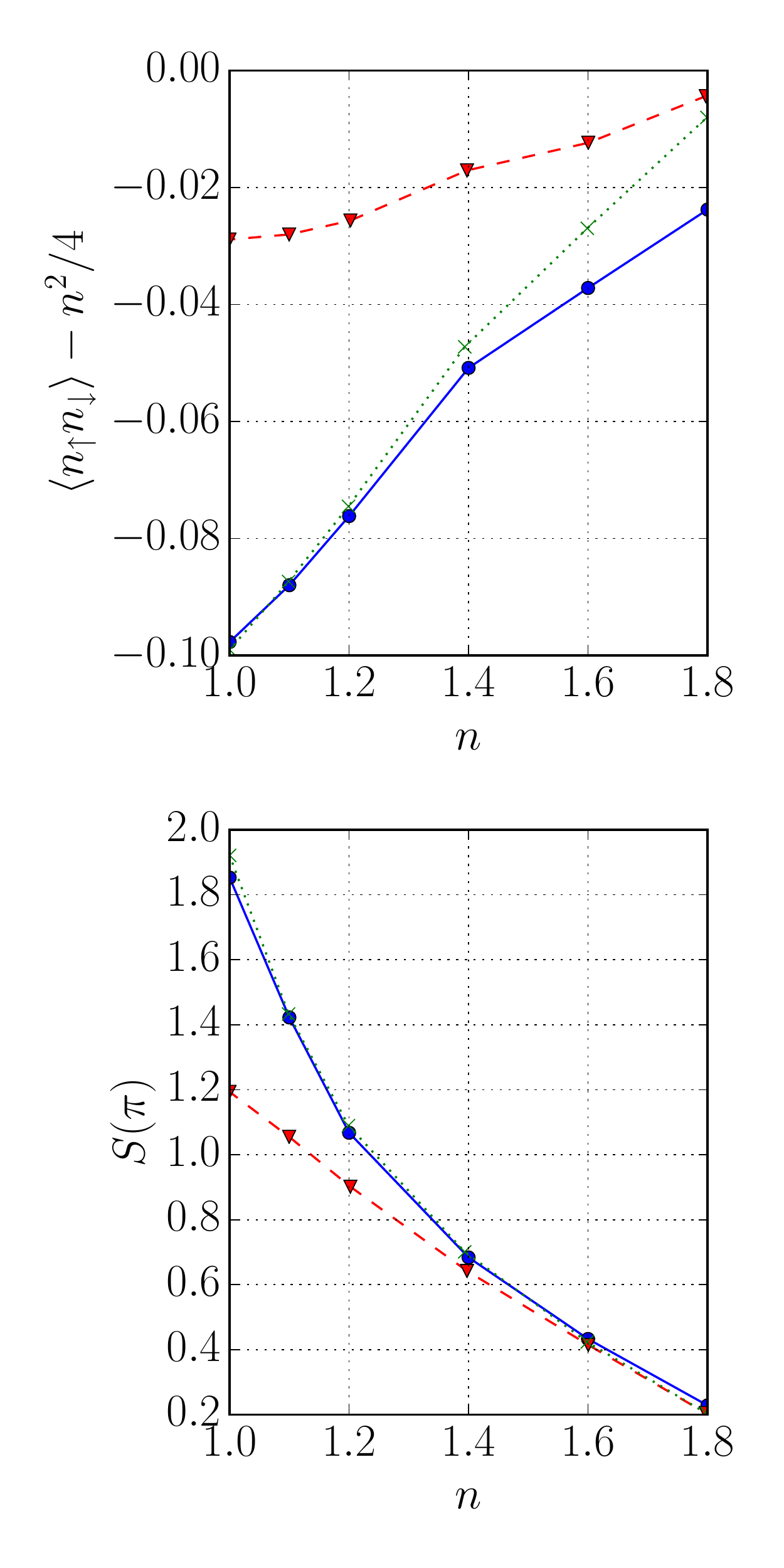} 
 \caption{QMC results for the full three-band model and the downfolded one-band model:  
 Comparison of the full model (circles), cRPA-screened model (triangles), and unscreened low-energy model (crosses) for $t^\prime=3$ and $\Delta=5$. The left panels show results for $U=4$ and the right panels for $U=8$.
 $n$ is the electron number in the target band.}
\label{fig:comparison}
\end{figure}

In Fig.~\ref{fig:comparison}, we compare the QMC results for the double occupation and spin-spin correlation function of the target band. The following three models are considered: (i) the original, full three-band model (blue),  which serves as the benchmark, (ii) the cRPA-screened one-band model (red), and (iii) the unscreened one-band model with bare interaction (red). The parameters are $t'=3$, $\Delta=5$, $U=4$ (left panels) and $U=8$ (right panels).  
We compare results for the same target-band filling, which is denoted by $n$, i.e. we use different chemical potentials in the three calculations. 
For our model, a  particle-hole transformation maps the data at $n\ge 1$  to those at $n\le 1$.

At half-filling, the system is in a Mott insulating state for $U>0$, while the solutions away from half-filling are metallic. 
The equal-time spin-spin correlation function is calculated in the band basis as 
\begin{equation}
S(q)   =   \frac{1}{N} \sum_{p,q, \sigma,\sigma'}  \sigma \sigma'  \Big\langle  c^{\dagger}_{\alpha,k,\sigma} c^{\phantom\dagger}_{\alpha,k+q ,\sigma} c^{\dagger}_{\alpha,p,\sigma'} c^{\phantom\dagger}_{\alpha,p-q ,\sigma'} \Big\rangle ,
\end{equation}
where $c^{\dagger}_{\alpha,k,\sigma} $  creates a  Bloch electron  in the target  band $\alpha$, with a $z$-component $\sigma$ of the spin. Fourier transformation of this quantity yields $ S(r) $,  from which we obtain the doublon density in the target band,
\begin{equation}
	\langle n_{\uparrow} n_{\downarrow} \rangle  =  \frac{ \langle n \rangle - S(r=0)  } {2},
\end{equation}
where $ \langle n \rangle $ is the filling of the target band. The spin  structure factor at wavevector $q=\pi$ is a measure of the antiferromagnetic correlations in the system, which are known to be enhanced by repulsive onsite interactions in the parameter regime considered here. Hence both, the suppression of the doublon density $\langle n_{\uparrow} n_{\downarrow} \rangle $ with respect to the non-interacting value $n^2/4$ as well as $S(\pi)$ are indicators for the strength of the effective onsite repulsion between the electrons in the respective models.  

In Fig.~\ref{fig:comparison} one can see that the cRPA-screened model (triangles) shows substantially smaller suppressions of the doublon density and enhancements of $S(\pi)$ than the unscreened model or the full model (circles). This implies that the cRPA-screening of the local repulsion is stronger than the screening effectively occuring in the full model. In fact, the unscreened model (crosses) that just leaves out the high-energy bands approximates the full model much better, as if there were almost no screening of the effective repulsion by the two high-energy bands, especially at half filling. Further ayway from $n=1$, both one-band approximations become insufficient to describe the doublon density, suggesting that interband effects are non-negligible in the large density limit of the target band. The comparison between the full model and the unscreened one-band model even indicates an anti-screening effect, which can by construction not be obained in cRPA. 

The data in Fig.~\ref{fig:comparison} demonstrate that the exact downfolding implicit in the full three-band QMC calculation corresponds to a much weaker screening than predicted by cRPA, or even to antiscreening. This finding is the first main result of this paper, which  motivates the cfRG study described in the following section.

\subsection{Insights from cfRG}
The cfRG method provides useful insights into why these discrepancies between the cRPA prediction and the full three-band model solution occur. First we study a two-band model with only one screening band above the Fermi level. Compared to the three-band case, this has less particle-hole symmetry and allows us to identify the processes that are missing in the cRPA calculation, and to quantify their effect. 

\subsubsection{Two-band model} 
\begin{figure}[b]
 \centering\includegraphics[width=.48\textwidth,clip]{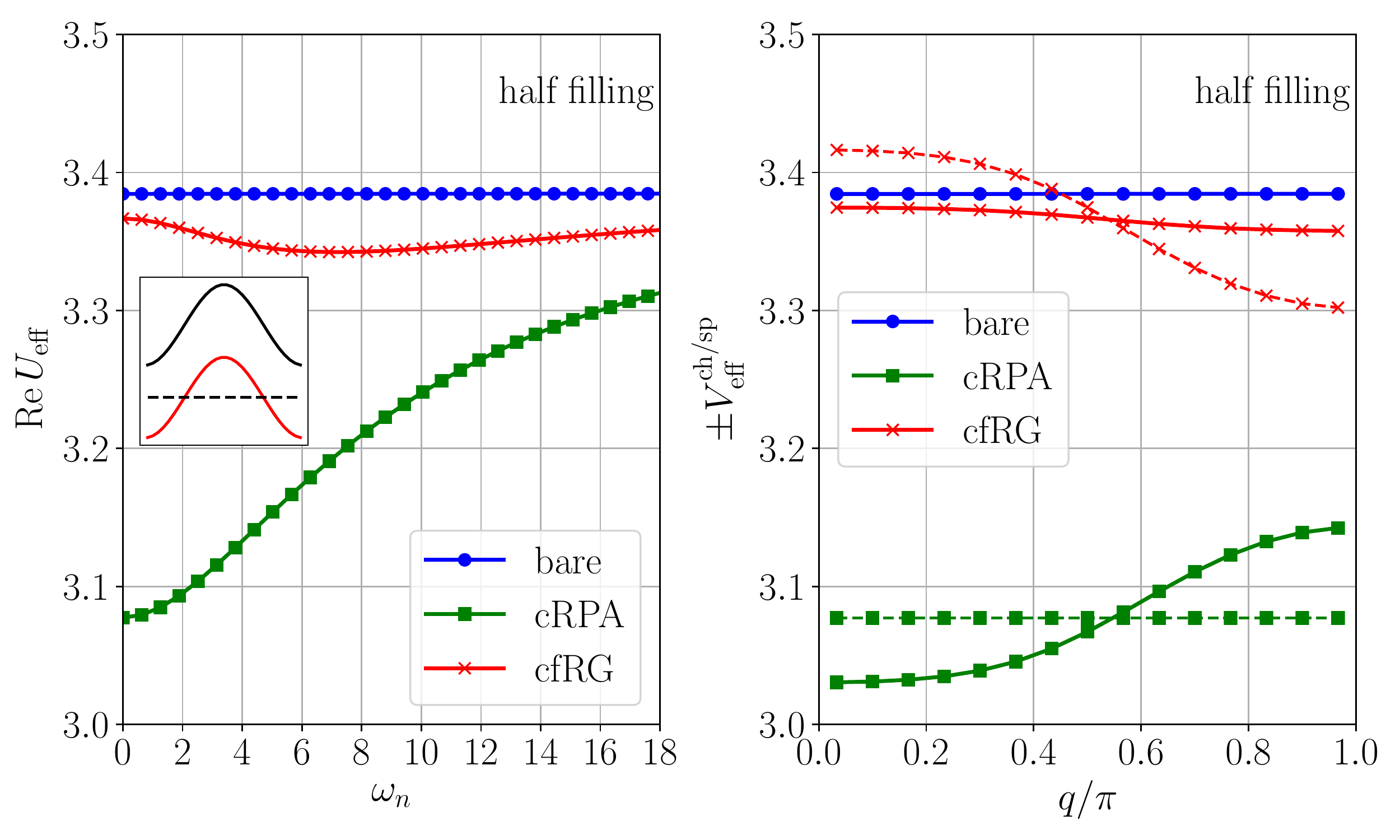}
 \caption{Data for the two-band model with $U=4$ and $\Delta=3$, $t'=1$, $T=0.1$ in  units of the hopping constant.  The `bare' curves are for the unscreened model, obtained by projecting the bare interactions of the two-band model onto the target band, without any loop corrections. The cRPA data take into account the corrections 
 by the cRPA diagram in Fig.~\ref{fig:diagrams}b). The cfRG eliminates the high-energy band taking into account the five  diagrams of Fig. \ref{fig:diagrams}b). 
Left panel: Effective target-band onsite charge interaction as a function of the transferred bosonic Matsubara frequency, 
 at half target-band filling. One observes a near-cancellation of corrections to the bare values in the cfRG and 10$\% $ screening in cRPA.
Right panel: Momentum dependence of the target-band charge (solid lines) and spin (dashed) interactions.  
 }
\label{fig:cancellation}
\end{figure}

In Fig.~\ref{fig:cancellation} we show the effective interactions generated by the cfRG downfolding 
for the two-band model with the bands as shown in the inset. 
The left plot presents data for the Matsubara-frequency dependent local repulsion $\text{Re}U_\text{eff}(i\omega_n)$ in the target band for three different cases: (i) the bare, unrenormalized interaction which is projected onto the target band (`unscreened model'), (ii) the cRPA interaction, and (iii) the cfRG interaction. 
Obviously, the cfRG and the bare target band interaction are quite close, while the cRPA produces substantial screening of about $10\%$ at low frequencies. Similarly, in the right panel of Fig. \ref{fig:cancellation}, we show the momentum dependence of the zero-frequency target-band charge (solid lines) and spin interaction (dashed lines), again for the three cases. The cfRG produces some wavevector-dependence or non-locality in the spin interaction, but hardly any such dependence in the charge channel. On average, the effective repulsion remains close to the bare one. In contrast to this, the cRPA displays different physics with stronger screening in both channels and more non-locality in the charge channel. Overall, the cfRG results are compatible with a near-cancellation of the loop corrections by the high-energy bands, which makes the screening much weaker than found in cRPA. 

The explanation of this behavior can be found by analyzing the right hand side of the cfRG equations for the target-band interaction, which is given by the diagrams in Fig.~\ref{fig:diagrams}b). These diagrams can be grouped into three channels: the diagrams cRPA, VC1 and VC2 driving together the coupling function $D_{o_1o_2}^{\Lambda}$ in Eq. \eqref{viobi}, the particle-particle (PP) channel renormalizing $P_{o_1o_2}^{\Lambda}$, and the crossed particle-hole (PHcr) channel renormalizing $C_{o_1o_2}^{\Lambda}$. If we insert the bare couplings into these diagrams and integrate out the screening bands completely, we obtain the second-order correction to the bare target-band interaction. These second-order terms contain the essence of the near-cancellation. 

\begin{figure}[t]
 \centering\includegraphics[width=.48\textwidth,clip]{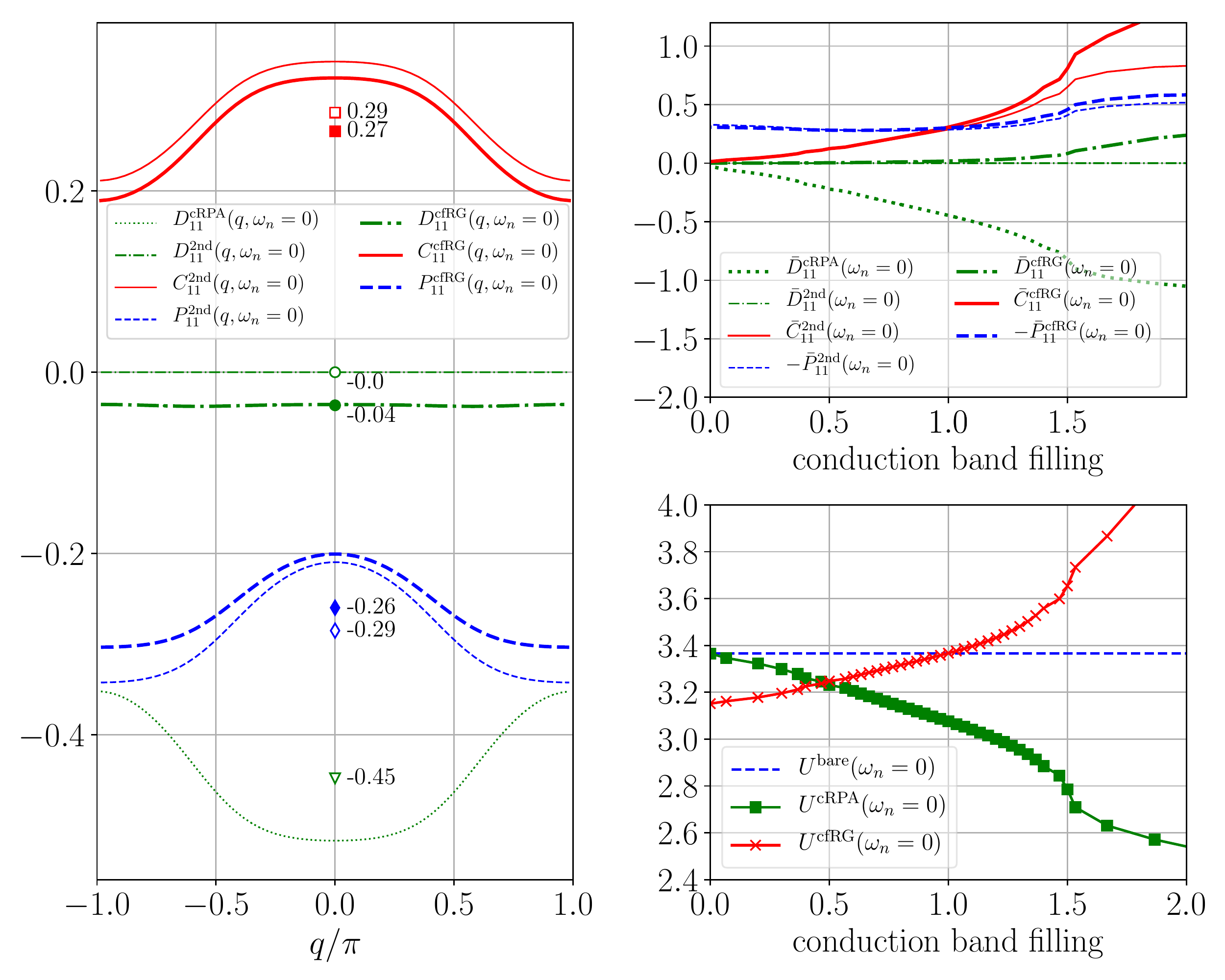}
 \caption{Data for the effective interaction in orbital 1 with the main weight in the target band after elimination of the high energy band of the two-band model of Fig.~\ref{fig:cancellation}. Left panel: Renormalized channel coupling functions $D^\Lambda_{11} (q)$, $C^\Lambda_{11} (q)$, and $P^\Lambda_{11} (q)$  as a function of the wavevector $q$ at zero Matsubara frequency $\omega_0$, with the bare interaction subtracted. The thick lines are for the full cfRG and the thin lines are for second order perturbation theory. While $D^\Lambda_{11} (q)$ is zero in the second order calculation, $C^\Lambda_{11} (q)$  and $P^\Lambda_{11} (q)$ cancel in the local interaction, i.e. when summed over the wavevector. The markers at $q=0$ (empty symbols for thin lines, filled symbols for thicker lines) annotated with numbers give the momentum-averages over the respective couplings.  
 The cancellation persists to a good approximation even in the full cfRG. 
 Right upper panel: Dependence of local, $q$-summed channel couplings on the electron density in the target band. At $n=1$ per site (spin-summed), $D^\Lambda_{11} (\omega_n=0)$, $D^\Lambda_{11} (\omega_n=0)$, and $D^\Lambda_{11} (\omega_n=0)$ cancel each other. The sign of $P^\Lambda_{11}(\omega_n=0)$ has been reversed to show the equality to $C^\Lambda_{11}(\omega_n=0)$ at half filling $n=1$.
 Right lower panel: Target-band filling dependence of the static onsite repulsion in the target band. 
 }
\label{fig:channelplot}
\end{figure}

The first crucial observation is that in this case, with the local intra-orbital repulsions chosen here, the screening contribution for $D_{o_1o_2}^{\Lambda}$ from the cRPA diagram (bottom left in Fig.~\ref{fig:diagrams}b)) is {\it cancelled} by the contributions from the other two diagrams, VC1 and VC2, in the direct particle-hole channel for any wavevector or frequency transfer through the diagrams. This can be seen in the left panel of Fig.~\ref{fig:channelplot}. There we plot the second-order correction to $D_{o_1o_2}^{\Lambda}$ as thin dashed-dotted line, which is exactly zero, together with the infinite-order cfRG result for the $D_{o_1o_2}^{\Lambda}$-correction, which is slightly negative, and the infinite-order cRPA correction without diagrams VC1 and VC2, which is about ten times more negative. We only show the diagonal contributions in orbital 1 that contributes the dominant weight to the target band after the elimination of the high-energy band (with the bare interaction subtracted). The target-band interactions in the three channels are obtained from these quantities by multiplying them with matrix elements from the orbital-to-band transformation for the incoming and outgoing legs. This is just a momentum-independent factor in our model that does not change the conclusions from this comparison.  

This cancellation of the `direct' particle-hole terms RPA, VC1 and VC2, is known from the single-band Hubbard model, where it is understood to arise from the momentum-independent bare interaction. It means, e.g., that the effective pairing interaction in second order can be computed just from the crossed particle-hole diagram. In the simple two- and three-band models with parallel bands and onsite-hybridization considered here, the cancellation still works, since we also have onsite interactions only and because the orbital-to-band transformation is momentum-independent. 
A closer analysis, which we checked numerically, shows that inter-orbital interactions lift this cancellation to some degree. 

The cancellation of the direct one-loop terms RPA, VC1 and VC2, means that in second order only the particle-particle and the crossed particle-hole diagram remain. These two however cancel as well upon integration over the transfer or total wavevector and frequency in the case of a 
half-filled conduction band. In Fig.~\ref{fig:channelplot} we show the wavevector dependence of the second-order corrections to $P_{o_1o_2}^{\Lambda}$ by the particle-particle and to $C_{o_1o_2}^{\Lambda}$ by the crossed particle-hole diagram. While the wavevector-dependence of the two terms is different, their averages (open diamond and open square) take opposite values, i.e. their sum cancels in the local target-band interaction (see the numbers $\pm 0.29$ for these averages in the plot). 
This cancellation is closely related to what occurs in one-band one-dimensional models, where it is responsible for the gaplessness of systems away from half filling with repulsive interactions despite the presence of logarithmically divergent particle-particle and particle-hole contributions to the perturbation series. 

As the diagrams RPA, VC1 and VC2 appear together on the right hand side of the flow equation for the coupling function $D^\Lambda_{o_1o_2}(q)$ in Eq.~(\ref{viobi}), this coupling function does not flow in second order. In higher orders, which are summed by the cfRG, the cancellation is violated by the generated momentum dependence of the couplings $D^\Lambda_{o_1o_2}(q)$, which will exhibit some mild flow.  In the left panel of Fig.~\ref{fig:channelplot} we also show the wavevector-dependence of the three interaction channels from Eq.~(\ref{viobi}). The cfRG only slightly changes the findings from the second-order analysis.

The next question is how general these effects are. In the lower right panel of Fig. \ref{fig:channelplot} we show the dependence of the target-band static onsite repulsion as a function of the conduction band filling. 
From this plot we see that the cancellation of the corrections to the target-band onsite repulsion in cfRG is only close to perfect at half filling, while below half filling we find a smaller value and above half filling a larger value. This trend can be easily understood from how the rigid band shifting with the filling changes the contributions of particle-hole and particle-particle one-loop diagrams. In those, due to the constraint of not treating screening solely due to the target band, the most important contributions come from the interband processes. For target band filling below half filling and with the high-energy band at positive band energies, we have more phase space for particle-particle processes. These act to suppress the local repulsion and we get a down-screening of the target-band interaction. For more than half filling, on the other hand, there are more possibilities for particle-hole processes with the high-energy band above. Then, the crossed particle-hole channel outweighs the particle-particle channel, and the effective interaction grows upon elimination of the high-energy band. Again, the cRPA shows a completely different trend.
Moving closer to complete filling of the target band, we find a strong particle-hole scattering between the target band top at the center of the Brillouin zone and the minimum of the high-energy band at the Brillouin zone edges. This significantly increases the spin- and charge target-band interaction at wavevector $q=\pi$, i.e. there is substantial antiscreening.

The upper right panel of Fig.~\ref{fig:channelplot} contains equivalent data for the filling dependence in the simple two-band model, split up into the three interaction channels and confirming the statements made above how the channel balance changes with target band filling.

Instead of changing the band filling, we can also destroy the balance between the particle-particle and the crossed particle-hole channel by deforming the band structure of the two bands. In the left panel of  Fig. \ref{fig:cfRGfig2} we show the filling dependence of the local target-band repulsion for the two-band model with a reversed hopping constant of the upper band. Now, the occupied and unoccupied parts of the target band are no longer symmetric at half band filling, and no cancellation is found at half filling. Similar effects can be obtained without changing the hopping but allowing for a local inter-orbital repulsion. We found that for $\Delta =5$, $U_1=U_2=4$ and inter-band repulsion $U'=2$, the target-band repulsion in cfRG grows by 7$\%$ compared to the unrenormalized value. 

\subsubsection{Three-band model}

\begin{figure}[ht]
 \centering\includegraphics[width=.48\textwidth,clip]{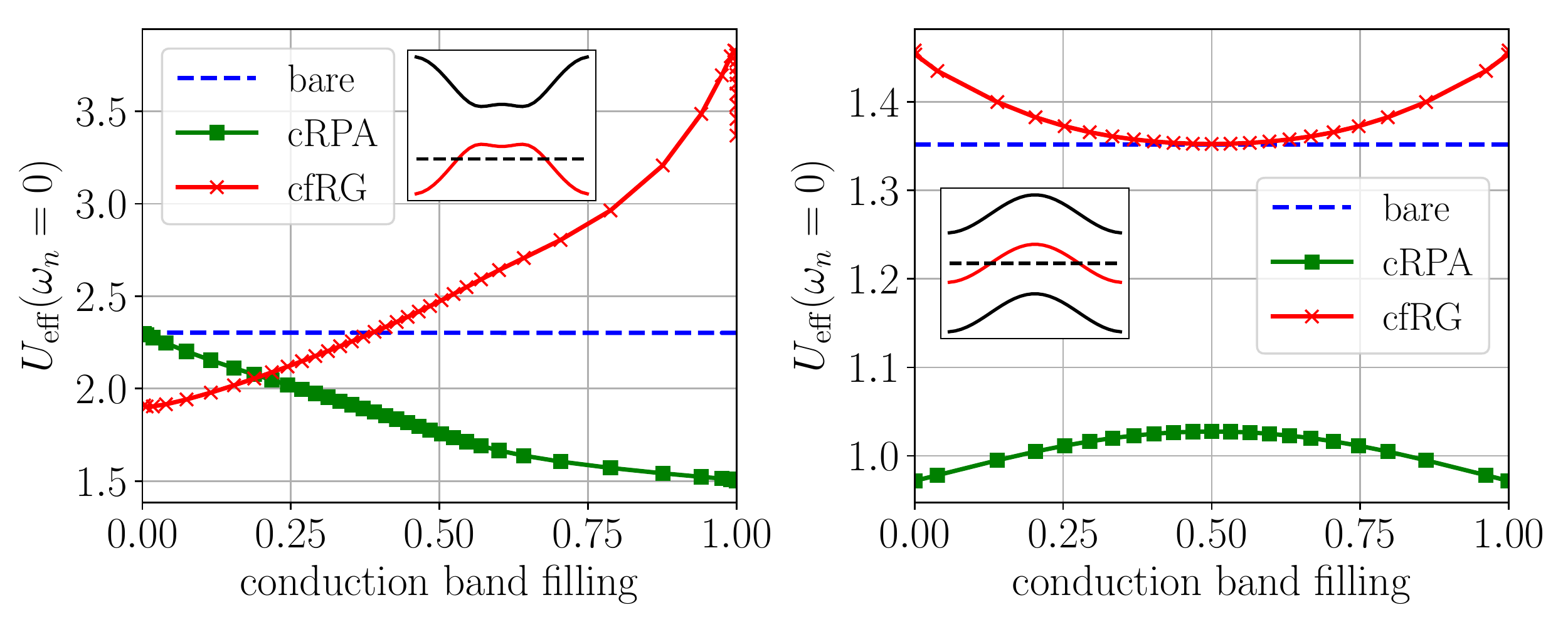}
 \caption{Left panel: Target-band filling dependence of the static onsite repulsion in the target band of the two-band model with inverted sign for the hopping of the upper band. The inset shows the non-interacting band structure, with the target band in red. Here, at half filling, the loop corrections do not cancel out and the cfRG antiscreens, while the cRPA screens the local repulsion down compared to the bare $U$ projected onto the target band. The strong variation near full band filling is due to the small dip at the top of the target band at $q=0$. 
Right panel: Target band filling dependence of the static onsite repulsion in the target band for the three-band model considered in the left panels of Fig.~\ref{fig:comparison}  ($\Delta=5$, $t'=3$ and $U=4$). Due to the symmetry of the bands, the filling dependence is also symmetric. At half filling, the cfRG loop corrections cancel almost perfectly: the cfRG value is about half a permille higher than the bare value. cfRG and cRPA predict opposite trends for the filling dependence.    
 }
\label{fig:cfRGfig2}
\end{figure}

The effective static interaction for the three-band case with equal hoppings and $U=4$ on the middle orbital (with the largest weight in the target band) is shown in the right panel of Fig.~\ref{fig:cfRGfig2}. Here, the bare target-band interaction of the unscreened model takes the value $1.35$ also found in Fig. \ref{fig:model1-cRPA}, while the cRPA interaction at half filling becomes $1.03$. The cancellation at half filling is perfect again for the effective static onsite repulsion. Now, as we have a symmetric band structure around the target band, the change of the local target-band repulsion comes out symmetric with respect to the half-filled target band. The cRPA screnning is substantial and increases away from half-filling, while the cfRG value deviates insignificantly from the bare value near half-filling. Note however that the momentum-dependent spin interaction at wavevector $q=\pi$ in the target band is mildly reduced by one percent. This goes in the right direction compared to the QMC data shown in the lower panels of Fig.~\ref{fig:comparison}, where the spin response at $q=\pi$ of the full three-band model is slightly weaker than the one in the unscreened model.

In contrast to cRPA, the cfRG result predicts anti-screening away from half filling. The QMC results for the doublon density in the upper panels of Fig.~\ref{fig:comparison} show that this antiscreening effect is indeed consistent with the full-model calculation and that the opposite trend predicted by cRPA is qualitatively wrong. The antiscreening does not show up in $S(\pi)$ in the lower panels, as $q=\pi$ is not the dominant scattering wavevector away from half filling.      

\section{Discussion}
\label{sec:discussion}

We have presented a model study for one-dimensional two- and three-band models in which only one band crosses the Fermi level. Using QMC, we computed the  equal time  spin response and the double occupation for three different cases: for the full multiband model, for the `unscreened' one-band model where the band away from the Fermi level is just ignored, and the cRPA-screened one-band model. The QMC data show that the double occupation and the spin correlations of the unscreened model solution are closer to the full multiband result than the cRPA one. Also, the QMC results imply antiscreening away from half-filling, which cannot be captured by the cRPA downfolding. 
The origin of the incorrect cRPA prediction could be traced back to cancellations between RPA and non-RPA diagrams, by systematic elimination of the high-energy bands using the more refined cfRG formalism. This scheme takes into account all five one-loop diagrams. For the given model, the combination of these diagrams leads to an {\it exact cancellation} of the loop corrections in second order in the bare couplings at half filing, and otherwise only to minor corrections that are quite different from the cRPA results. 

By exploring the cfRG interactions for other band structures and for different target band filling we showed that the cancellation of the loop corrections only occurs under special circumstances but that cfRG and cRPA lead in general to quantitatively and also qualitatively different predictions for the effective interaction. Most notably, the cfRG can also predict  antiscreening, i.e. result in larger effective interactions, while the cRPA always screens the interaction. Such an antiscreening effect is indeed found in our exact QMC benchmark calculation away from half-filling.  

Additional systematic studies have to be performed to assess whether the failures of the cRPA method revealed by these examples are generic or represent pathological situations. The results are consistent with the findings of a previous EDMFT investigation of downfolding in a three-dimensional three-band Hubbard model~\cite{Shinaoka2015}. They also look consistent with the failures of partial diagrammatic summations demonstrated for a Hubbard model in Ref.~\onlinecite{Gukelberger2015}. From other cfRG works, it is known that the cfRG-corrections to cRPA vanish if the high-energy bands have a different symmetry than the conduction or target bands \cite{Ginza2015}. Such a situation occurs, e.g., in monolayer graphene. Furthermore, in Ref. \onlinecite{Honerkamp2018}, non-local bare interactions were considered. For those, the screening differences between cRPA and cfRG became small in the long-range limit, or for $q \to 0$. Nevertheless, the effective interaction at nearest neighbors on the lattice showed clear differences between cRPA and cfRG. 

In view of more substantial studies which are closer to ab-initio modeling of materials, it is important to also comment on the numerical effort of the methods. cRPA has been integrated into the standard first-principles framework and has also been used in many recent materials studies \cite{Miyake2009,Tomczak2009,Vaugier2012,Werner2012,Werner2015}. In contrast, the cfRG method has up to now been confined to model studies of few-band systems. The main reason is that the effective interaction depends in general on three wavevectors and three frequencies. This prevents the application to models with more than two or three bands, at least without resorting to more approximations. Recently however, various physically appealing approximation schemes have been tested. They significally reduce the numerical load. Most of these development are summarized in a recent preprint \cite{Honerkamp2018}, where also efficient and meaningful approximations for the orbital content of the interactions are presented and tested. With the techniques advocated there, the numerical effort of the cfRG methods should be comparable to cRPA. 

\acknowledgements

HS, PW  and FFA acknowledge support from the Deutsche Forschungsgemeinschaft (DFG) via FOR 1346 and HS and PW  from the Swiss National Science Foundation via Grant 200021-165539.    CH acknowledges funding via DFG Ho2422/10-1, HO2422/11-1 and the DFG research training group 1995 Quantum Many-Body Methods in Condensed Matter Systems€. Michael Kinza and Johannes Hofmann are acknowledged for collaborations in early stages of the project.

\end{document}